\documentclass[a4paper,12pt]{article}
\usepackage[utf8]{inputenc}

\usepackage{amsfonts}
\usepackage{amsmath}
\usepackage{amssymb}
\usepackage{mathtools}
\usepackage{amsthm}
\usepackage{thmtools}
\usepackage{bm}
\usepackage{authblk}\bigskip

\usepackage{xcolor}
\usepackage{enumitem}
\usepackage{hyperref}

\usepackage[bb=dsserif]{mathalpha}
\usepackage{tikz}
\usepackage{tikz-3dplot}
\usetikzlibrary{calc}
\usetikzlibrary{perspective}

\usepackage[backend=biber,style=numeric]{biblatex}
\newcommand{\mcal}[1]{\mathcal{#1}}

\newcommand{\powerset}[1]{\mcal{P}_{#1}}

\newcommand{\MSN}{\frac{s!(n-s-1)!}{n!}}

\DeclarePairedDelimiterX{\innerprod}[2]{\langle}{\rangle}{#1, #2}

\DeclareMathOperator{\Ima}{Im}

\DeclareMathOperator{\Ker}{Ker}

\newcommand{\C}{\chi} %

\newcommand{\NS}{N \setminus S}
\newcommand{\NT}{N \setminus T}

\newcommand{\sumSN}{\sum_{S \subseteq N}}

\newcommand{\sumSNminusi}{\sum_{S \subseteq N \setminus \{i\}}}

\newcommand{\fractwoN}{\frac{1}{2^n}}

\newcommand{\fractwoNminusone}{\frac{1}{2^{n-1}}}
\newcommand{\fractwo}{\frac{1}{2}}
\newcommand{\Cprime}{\C'}

\newcommand{\bigbig}[1]{\big(#1\big)}

\newcommand{\gameN}{\mathcal{G}^{N}}
\newcommand{\GammaN}{\Gamma^{N}}

\newcommand\restrict[1]{\raisebox{-.5ex}{$|$}_{#1}}

\newcommand{\elledueV}{\mathit{l}^2(V)}
\newcommand{\elledueE}{\mathit{l}^2(E)}

\newcommand{\ddstar}{\mathrm{d}^*}
\newcommand{\ddiff}{\mathrm{d}}

\newcommand{\dds}[1]{\mathrm{d}_{#1}}

\newcommand{\R}{\ensuremath{\mathbb{R}}}				%

\newcommand\restr[2]{{%
  \left.\kern-\nulldelimiterspace %
  #1 %
  \vphantom{\big|} %
  \right|_{#2} %
  }}

\declaretheoremstyle[
spaceabove=6pt, spacebelow=6pt,
headfont=\normalfont\bfseries,
notefont=\mdseries\bfseries, notebraces={(}{)},
bodyfont=\normalfont,
]{normalstyle}

\declaretheoremstyle[
spaceabove=6pt, spacebelow=6pt,
headfont=\normalfont\bfseries,
notefont=\mdseries\bfseries, notebraces={(}{)},
bodyfont=\itshape,
]{plainthm}

\declaretheorem[numberwithin=section,
name=Theorem,
refname={theorem,theorems},
Refname={Theorem,Theorems},
style=plainthm]
{thm}

\declaretheorem[sibling=thm,name=Definition,
refname={definition,definitions},
Refname={Definition,Definitions},
style=plainthm
]
{defin}

\declaretheorem{lem}[
name=Lemma,
sibling=thm,
style=plainthm
]
\declaretheorem{prop}[
name=Proposition,
sibling=thm,
style=plainthm
]
\declaretheorem{cor}[
name=Corollary,
sibling=thm,
style=plainthm
]

\declaretheorem{notaz}[
name=Notation,
sibling=thm,
style=normalstyle
]
\declaretheorem{rmk}[
name=Remark,
sibling=thm,
style=normalstyle
]

\declaretheorem{ex}[
name=Example,
sibling=thm,
style=normalstyle
]

\providecommand{\MSC}[1]
{
  \small	
  \textbf{\textit{MSC 2020:} } #1
}

\makeatletter
\newcommand*\rel@kern[1]{\kern#1\dimexpr\macc@kerna}
\newcommand*\widebar[1]{%
  \begingroup
  \def\mathaccent##1##2{%
    \rel@kern{0.8}%
    \overline{\rel@kern{-0.8}\macc@nucleus\rel@kern{0.2}}%
    \rel@kern{-0.2}%
  }%
  \macc@depth\@ne
  \let\math@bgroup\@empty \let\math@egroup\macc@set@skewchar
  \mathsurround\z@ \frozen@everymath{\mathgroup\macc@group\relax}%
  \macc@set@skewchar\relax
  \let\mathaccentV\macc@nested@a
  \macc@nested@a\relax111{#1}%
  \endgroup
}
\makeatother
\title{The Shapley-Hodge Associated Game}
\author[1]{Antonio Mastropietro}
\author[2]{Francesco Vaccarino}
\affil[1]{Department of Data Science, EURECOM, France, antonio.mastropietro@eurecom.fr}
\affil[2]{Department of Mathematical Sciences, Politecnico di Torino, Italy, francesco.vaccarino@polito.it
}
\DeclareUnicodeCharacter{0301}{*************************************}

\begin{document}

\maketitle
\abstract{In cooperative game theory, associated games allow providing meaningful characterizations of solution concepts. 
Moreover, generalized values allow computing an influence or power index of each coalition in a game.
In this paper, we view associated games through the lens of game maps and we define the novel Shapley-Hodge Game, briefly ``SHoGa''.
We characterize SHoGa via an axiomatic approach as a generalized value, and we thoroughly discuss the consistency properties of its associated game.
Furthermore, we describe the Hodge decomposition of an oriented graph representing the transitive closure of the Hasse diagram of coalitions in the game.
Finally, we show how SHoGa is linked to the solution of the Poisson equation derived from such a decomposition.
}

\vspace{0.3cm}

\MSC{91A12, 05C50, 15A24}

\section{Introduction}

In cooperative game theory, associated games have assumed a significant role in the study of cooperation structure and worth allocation to players.
An associated game is nothing else than the results of a transformation providing a new view of the game being studied \cite{hart1989potential, hamiache2001associated}.
It is known that relating and comparing properties of games and associated games allows for deriving novel characterization of solution concepts, particularly values \cite{hamiache2001associated, hamiache2020associated, kleinberg2018note_associated_consistency}.
However, the past literature has mainly focused on sharing rules by themselves neglecting the functional relationship between games.
Hence, we propose to study game maps so that an associated game results in the image game through a game map.
The idea of game maps is not completely novel in the literature and it has been studied to characterize values in terms of associated consistency properties \cite{kleinberg2018note_associated_consistency}. 
In other works, the term generalized value has been used to describe the functions aiming at computing a power or influence index of coalitions in a game, generalizing the concept of value for players \cite{marichal2000influence, marichal2007axiomatic, flores2019evaluating}.
However, to the best of our knowledge, the linkage between the two perspectives seems not to be extensively investigated before.
Unsurprisingly, game maps allow revisiting a number of past works on well-studied values \cite{hart1989potential, hamiache2001associated, oishi2016duality_game, hamiache2020associated, hwang2006associated, xu2013axiomatization}. 
Indeed, we illustrate that many characterizations of values can be viewed in the light of studying the properties of game maps. 
In addition, we show that the axiomatic approach, inherited from generalized values, can be transferred to the study of game maps and to associated games.
In other words, we discuss a significant set of properties or axioms that game maps should have in a similar fashion to generalized values. 
Note that while axioms for values usually involve properties required on players, like the dummy player or symmetric axioms of the Shapley value \cite{1953_Shapley_value}, the axiomatic approach for game maps involves properties of coalitions.
Furthermore, inspired by the first, classical characterization of the Shapley value \cite{1953_Shapley_value} we isolate a set of properties that uniquely characterizes a noteworthy game map.
Fixed a set of players $N$, we focus on the linear space of transferable utility games (TU-games) on $N$, hereafter denoted by $\gameN$. 
In particular, we propose a game map $\C:\gameN\to \gameN$ which computes the newly defined Shapley-Hodge associated game $\C_u$ as:
\begin{equation}\label{eq:shapley_hodge_associated_game}
    \C_u(S) = \fractwo \bigbig{u(N) - u(N \setminus S) + u(S)} \quad \forall u \in \gameN.
\end{equation}
Then, we show that strong relationships exist between probabilistic values \cite{weber1977probabilistic} and probabilistic generalized values \cite{marichal2007axiomatic} applied to $u \in \gameN$ and $\C_u$.
In addition, we study the link between $\C_u$ and families of games of various interests, like bilateral, constant sum, cohesive, and superadditive games.
Moreover, we describe $\C_u$ in terms of the dual game \cite{ruiz1998family, oishi2016duality_game, grabisch2016set} and the quotient game \cite{aumann1974cooperative}.
Interestingly, the $\C_u$ associated game appears to follow the equal division of surplus principle \cite{van2009axiomatizations, ramon2022equal} in the quotient game \cite{owen1977values} of two complementary coalitions.

Finally, we investigate the properties of the linear map $\C$ in the vector space of TU-games.
The recent work of \cite{STERN2019186} derived a characterization of the Shapley value in terms of the combinatorial Hodge decomposition of an oriented hypercube graph describing the marginal contributions of players in the game.
Inspired by their work, we provide a characterization of the associated game $\C_u$ using the same Hodge decomposition of the transitive closure of the Hasse diagram graph describing the inclusion relation of all the game's coalitions. 
We conclude with some examples and remarks about the novel game map $\C$ of eq. \eqref{eq:shapley_hodge_associated_game}.

\section{Axiomatic characterization}
\subsection{Preliminaries}
A Transferable-Utility cooperative game (TU-game) is a pair $(N, u)$: $N$ is a set of players of cardinality $n \in \mathbb{N}$, and $u \colon \powerset{N} \to \R$ is such that $u(\emptyset) = 0$, where $\powerset{N}$ is the power set of $N$.
In the following, the term game refers to a TU-game $(N, u)$, and it is identified with its characteristic function $u$, while $N$ is assumed to be $\{1, \dots, n\}$.
A player participating in the game is represented as an element $i\in N$, whereas each $S \subseteq N$ denotes a coalition of players.
We denote by $\NS$ the complementary coalition.
The function $u$ is called the characteristic function of the game, and $u(S)$ expresses the worth of coalition $S$ in the cooperation of the players. 
A coalition $S \subseteq N$ for a game $u\in \gameN$ is called bilateral if $u(S) = u(\NS)$, as studied by \cite{mlodak2007bilateral_constant_sum_games}. 
A game is bilateral if all its coalitions are bilateral.
We define $S$ as a null coalition if $u(S \cup T) = u(T)$ for $\forall T \subseteq \NS$.

If $S \subseteq N$, a subgame $(S, u\restrict{S})$ of a game $(N, u)$ is a TU-game where the characteristic function is defined on each $T \in \powerset{S}$ as $u\restrict{S}(T) = u(T)$. 
A game $u$ is superadditive if $u(S \cup T) \geq u(S) + u(T)$ whenever $S \cap T = \emptyset$.
A game $u$ is cohesive if for all partitions $\mathfrak{P} = \{S_1, \dots, S_k\}$ of $N$, $u(N) \geq \sum_{j=1}^k S_j$.
Note that a superadditive game is cohesive.
In addition, $u$ is constant sum if for each coalition $S$, $u(S) + u(\NS) = u(N)$.
It is known \cite{1994_Osborne_CourseGameTheory} that the set of TU-games, indicated by $\gameN$, is a vector space with obvious functional operations. 

Among the games in $\GammaN$ are of paramount relevance the unanimity games $\theta_S$ given by $\theta_S(T)=1$ if and only if $S\subseteq T$, for all $S, T \subseteq N$.
The set of unanimity games $\{\theta_S\}_{S\subseteq N}$ is a linear basis of $\gameN$ (as real vector space) see \cite{1994_Osborne_CourseGameTheory}. 

Given a game $u$, its zero-normalization is $u_z(S) = u(S) - \sum_{i \in S} u(i)$ \cite{weber1977probabilistic};
the dual-game and the anti-dual game are defined as $u^*(S) = u(N) - u(\NS)$ and $-u^*(S)$, respectively \cite{oishi2016duality_game}.
In a cooperative game, it is possible that sets of players behave as blocks. 
The quotient game defined below represents the situation of a fixed coalitional structure where a partition of $N$ into coalitions acts as the set of players \cite{aumann1974cooperative}.
Consider a game $u \in \gameN$ with $n \geq 2$, and $K = \{1, \dots, k\}$ such that $k\leq n$. Let $\mathfrak{P} =\{S_1, \dots, S_k\}$ be a partition of $N$. The quotient game $u^{\mathfrak{P}} \in \mathcal{G}^K$ is such that the set of players consists of the coalitions in $\mathfrak{P}$ and for each $\{i_1, \dots, i_l\} \subseteq K$, $u^{\mathfrak{P}}(\{i_1, \dots, i_l\}) =  u(\cup_{j=1}^l S_{i_j})$.

For a $u\in \gameN$, a payoff vector for $N$ is a vector in $\R^{n}$ allocating gain to each player $i\in N$. 
For a set of such games, a solution is a mapping that associates a set of payoff vectors to each games.
An axiom is a property that is desirable for a solution to satisfy. 
A value is a solution on a domain of games associating a unique payoff vector to each game.
A characterization of a value is a set of properties on the domain of games over which it is defined that allows deriving a uniqueness theorem among the values on the same domain. 
The value determines a rule to allocate the worth of the grand coalition among the individual players. 
A value $\psi$ is probabilistic \cite{weber1977probabilistic} if, for each player $i$, there exist a family of constants $p_S^i(N) \in \R$ for each coalition $S \subseteq N \setminus i$, such that: 
\begin{equation*}
    \psi_i(u) = \sum_{S\subseteq N \setminus i} p_S^i(N) \, \bigbig{u(S \cup i) - u(S)}, \quad \text{ with } \sum_{S\subseteq N \setminus i} \, p_S^i(N) = 1.
\end{equation*}
Note that a value satisfying the linearity and dummy player axioms is probabilistic \cite{grabisch1999axiomatic}.
Examples of probabilistic values are the Shapley value and the Banzhaf value \cite{banzhaf1964weighted}. 
In particular, we will focus on the Shapley value, which is probably the most famous and studied value on TU-games.
Given a game $u \in \gameN$, for each player $i$ the Shapley value is defined as:
\begin{equation}\label{eq:def_shapley_value}
    \phi_i(u) = \sum_{S\subseteq N \setminus i} \MSN (u(S \cup i) - u(S)).
\end{equation}
Many axiomatic characterizations have been shown in the literature \cite{algaba2019handbook}.
Among the others, we mention the first characterization by Shapley himself \cite{1953_Shapley_value}, and the one using associated consistency by \cite{hamiache2001associated}.

A significant extension of probabilistic values are the probabilistic generalized values \cite{marichal2000influence, marichal2007axiomatic, flores2019evaluating}, 
defined as 
\begin{equation}\label{eq:probabilistic_generalized_values}
    \psi_S(u) = \sum_{T\subseteq N \setminus S} p_T^S(N) \, \bigbig{u(T \cup S) - u(T)} , \quad \text{ with } \sum_{S\subseteq N \setminus i} \, p_T^S(N) = 1.
\end{equation}
The idea of generalized values is to measure a coalition's power, strength, or influence in a game \cite{marichal2000influence}.
Again, examples are the Shapley or Banzhaf generalized values \cite{marichal2007axiomatic}.
Recall that a probabilistic generalized value satisfies group rationality if, for all superadditive game $u \in \gameN$, $\psi_S(u) \geq u(S)$ \cite{flores2019evaluating}.

\subsection{The unifying definition: game map}
In this section, we define and study functions defined on $\gameN$, hereby called game maps.

\begin{defin}\label{defin:game_map}
A game map is a function $\gamma$ that assigns to each game $u \in \gameN$ the associated game $\gamma_u \in \gameN$ on the same set of players. 
We denote by $\GammaN$ the set of game maps from $\gameN$ to itself. 
\end{defin}
Note that game maps have been already defined in the literature with the term ``generalized value'' \cite{marichal2007axiomatic, flores2019evaluating}. 
However, to the best of our knowledge, the link between associated games and game maps has not been studied before.

Observe that if $\gamma, \gamma'$ are game maps, then $c \, \gamma$ is a game map for all $c \in \R{}$, and $\gamma + \gamma'$ is a game map as well.
Furthermore, note that the function $\C$ of Equation \ref{eq:shapley_hodge_associated_game} is a game map. 
Indeed, (i) $\C_u(\emptyset) = 0, \, \forall u \in \gameN$ and (ii) it associates $\C_u\in \gameN$ to each $u \in \gameN$.

We can use game maps to review significant results in the literature.
First of all, the potential defined in \cite{hart1989potential} as
\begin{equation*}
    P(N, u) = \sumSN \frac{(s-1)! (n-s)!}{n!} u(S)
\end{equation*}
induces the game map $\rho$ with associated game $\rho_u(S) = P(S, u\restrict{S})$. 
The potential differential $D^{i}P(N, u)$ of $u$ to a player $i$ is easily derived from $\rho$:
\begin{equation*}
    D^{i}P(N, u) = P(N, u) - P(N \setminus \{i\}, u\restrict{N \setminus \{i\}}) = \rho(N) - \rho(N \setminus \{i\}).
\end{equation*}
Proceeding further, the family of associated games with parameter $t \in \R{}$ defined by:
\begin{equation}\label{eq:hamiache_associated_game}
    \eta_u^{t}(S) = u(S) + t \,\sum_{j \in N \setminus S} \bigbig{u(S \cup \{j\}) - u(S) - u(\{j\})}, \quad \forall S \subseteq N
\end{equation} 
defined in \cite{hamiache2001associated} are image games through suited game maps. 
In analogy to their characterization of the Shapley value in terms of associated consistency, we provide the following definition regarding game maps and values, already mentioned in \cite{kleinberg2018note_associated_consistency}. 
\begin{defin}\label{def:associated_consistency_game_map}
    A value $\psi_i$ on the domain $\gameN$ is associated consistent with $\gamma \in \GammaN$ if and only if $\psi_i(u) = \psi_i(\gamma_u)$ for each $i \in N$.
\end{defin}
It's worth noting that \cite{hwang2006associated} defines another associated game to characterize the value called equal allocation of non-separable cost (EANS) of \cite{moulin1985separability}.
In addition, \cite{hamiache2020associated} studies games with a cooperation structure through a family of associated games similar to the one of eq. \eqref{eq:hamiache_associated_game}.

Moreover, we note that the function $\kappa$ deriving the Harsanyi dividends of a game \cite{harsanyi1963simplified} is a game map:
\begin{equation*}
    \kappa_u(S) = \sum_{T\subseteq S} (-1)^{s-t} \, u(T), \quad \forall u\in \gameN.
\end{equation*}
In addition, the synergy $\omega$ of a game is a game map as well \cite{grabisch1999axiomatic}:
\begin{equation*}
    \omega_u(S) = \sum_{T\subseteq S} \, u(T) , \quad \forall u\in \gameN.
\end{equation*}
Furthermore, $\omega$ is actually the inverse of $\kappa$ in the sense that $\omega_{\kappa_u} = \kappa_{\omega_u} = u$, as can be shown by the properties of the M\"{o}ebius transform \cite{grabisch2016set}.

We observe also that a probabilistic generalized value is a game map.
Indeed, from eq. \eqref{eq:probabilistic_generalized_values} it holds 
that $\psi_{\emptyset}(u) = 0$ for each $u \in \gameN$.
Vice versa, it is easy to show that $\C_u$ is a probabilistic generalized value as well, with \begin{equation*}
   p_T^S(N) = \begin{cases}
       \frac{1}{2} \text{ if } T \in \{\emptyset, \NS\} \\
       0 \text{ otherwise}.
   \end{cases}
\end{equation*}

As last examples, we observe that the zero-normalization, dual and anti-dual games are game maps. 
In particular, $\C$ results in (see eq.(\ref{eq:shapley_hodge_associated_game}))
\begin{equation}\label{eq:chi_wrt_dualgame}
\C_u = \frac{1}{2}\bigbig{u^* + u}.
\end{equation}
The above equation suggests that $\C$ can be interpreted as the average of the worth $S$ can obtain by itself and the worth that $\NS$ cannot prevent $S$ to obtain in $u$ \cite{oishi2016duality_game}.
\begin{defin}
The Shapley-Hodge game map $\C \in \GammaN$ is the game map sending every $u \in \gameN$ to $\C_u$, its Shapley-Hodge associated Game, briefly ``SHoGa''.
\end{defin}
In the following sections, we will provide motivations for linking the $\C$ function with the Shapley value and Hodge theory.

\subsection{Axiomatic approach to game maps}

We can now show a characterization of the game map $\C$.
\begin{defin}\label{defin:axioms_game_map}
$\gamma \in \GammaN$ satisfies one of the following axioms if the corresponding statement holds:
\begin{itemize}[labelwidth=1.2cm, align=parleft,leftmargin=1.4cm]
    \item[AvEFF] Average Efficiency: $\forall u \in \gameN$, $
    \sumSN \gamma_u(S) = 2^{n-1} u(N)$.
    \item[NLL] Null Coalition: $\forall u \in \gameN$, if $S \subseteq N$ is a null coalition for $u$, then $\gamma_u(S) = 0$.
    \item[BLT] Bilaterality: $\forall u \in \gameN$, if $S \subseteq N$ is bilateral for $u$, then $S$ is bilateral for $\gamma_u$ as well. 
    \item[CS] Constant sum: $\forall u \in \gameN \quad \gamma_u(S) + \gamma_u(\NS) = \gamma_u(N)$
    \item[LIN] Linearity: $\forall u, v \in \gameN, \, \forall a, b \in \R, \quad \gamma_{a u + b v } = a\gamma_u + b \gamma_{v}$.
\end{itemize}
\end{defin}
The AvEFF axiom requires that the average of the associated game worths should be equal to the worth of the grand coalition. 
The averaging factor $\frac{1}{2^{n-1}}$ is the number complementary pairs $(S, \NS)$ participating in the game.
The NLL axiom extends the axiom for players in the context of coalitions.
This axiom is linked to the dummy coalition axiom for probabilistic generalized values introduced by \cite{marichal2007axiomatic}.
The BLT axiom expresses an equal mapping for complementary coalitions having the same original worth.
The above two axioms can be explained as a coalitional version of a fairness principle, involving pairs of complementary coalitions.
This view suggests that the associated game $\gamma_u$ could be interpreted as a fair version of $u$.  
The CS axiom is a requirement on complementary pairs of the associated game: each pair should sum to the associated worth of the grand coalition.
Finally, the LIN axiom is inspired by the linearity properties of values.
We can now show that the game map $\C$ is characterized by the listed properties.
\begin{thm}\label{thm: chi_axiom}
Assume that $\gamma \in \GammaN$ satisfies EFF, LIN, NLL, CS, BLT.
Then $\gamma_u = \C_u$ for all $u\in \gameN$. Therefore $\C$ is the unique game map satisfying EFF, LIN, NLL, CS, BLT.
\end{thm}
\begin{proof}
Since $\gamma$ satisfies LIN, it is a linear endomorphism of $\gameN$. Therefore, it is enough to prove the statement on a basis of $\gameN$. We opt for $\{\theta_S\}_{S\subseteq N}$, where $\theta_S$ are the unanimity games on $N$.%

Fix $S\subseteq N$ and consider $T\subseteq N$, there are only three cases: 

\begin{enumerate}
    \item[(i)] $T\cap S = \varnothing$. Let $R\subseteq N$ be such that $T\cap R =\varnothing$, then $S\subseteq T\cup R$ if and only if $S\subseteq R$ so that $T$ is null and, therefore, $\gamma_{\theta_S}(T)=0$. 
    \item[(ii)] $T\cap S = S$ i.e. $S\subseteq T$. In this case $(\NT)\cap S=\varnothing$, therefore $\gamma_{\theta_S}(T)=\gamma_{\theta_S}(N)$ by $CS_S$ and case (i).
    \item[(iii)] $\varnothing \neq T\cap S \neq S$. In this case also $\varnothing \neq (\NT)\cap S \neq S$, therefore $\theta_S(T)=\theta_S(\NT)=0$, hence $\gamma_{\theta_S}(T)=\gamma_{\theta_S}(\NT)=\frac{1}{2}\gamma_S(N)$ by $BLT_S$ and $CS_S$.
\end{enumerate}
The result follows thanks to the Lemma below.
\end{proof}

\begin{lem}
Assume that $\gamma \in \GammaN$ satisfies $AvEFF$ and $CS$. 
Then $\gamma_u(N) = u(N)$, for all $u\in \gameN$.
\end{lem}
\begin{proof}
$CS$ and $AvEFF$ imply that 
\[\gamma_u(N)=\sum_{S\subseteq N}(\gamma_u(S)+\gamma_u(\NS))=2u(N).\]
\end{proof}

\begin{prop}
    The kernel of $\C$ is 
    \begin{equation}\label{eq: kerc}
        \ker \C = \{u\in\gameN\,:\, u(A)=u(N\setminus A), \,\forall A\in\powerset{N}\}
    \end{equation}
    In particular, if $u\in \ker \C$, then $u(N)=u(\emptyset)=0$.
\end{prop}
\begin{proof}
 First, $\C_u(N)=0$ if and only if $u(N)=0$. The result follows by applying the definition of $\C_u(S)$ for $S\in\powerset{N}$ with $u(N)=0$.   
\end{proof}

\section{Properties of SHoGa}
We may now investigate the properties of SHoGa in terms of values and interesting classes of games.
First, we slightly extend a result already known for the Shapley value \cite{driessen1991survey}.
\begin{prop}\label{prop:shapley_value_formula_complementary}
If $\psi$ is a probabilistic value then 
\begin{equation}\label{eqref:formula_complementary_linear_value}
    \psi_i(u) = \sumSNminusi p_{S}^i \bigbig{u(N \setminus S) - u(S)}
\end{equation}
\end{prop}
\begin{proof}
Setting $R = N\setminus (S \cup i)$, from the definition of probabilistic value it follows that: 
\begin{align*}
    \psi_i(u) &= \sum_{R \subseteq N\setminus \{i\} }  p_{R}^i \, \bigbig{u(N \setminus R) - u(N \setminus (R \cup i))\,} \\ 
    &= \sumSNminusi p_{S}^i \, \bigbig{u(N \setminus S) - u(S)}
\end{align*}
\end{proof}

\begin{prop}\label{prop:phi_i_of_chi_shapley}
If $\psi$ is a probabilistic value, then:
\begin{equation}\label{eq:shapley_of_chi_S}
    \psi_i(\C_u) = \psi_i(u)
\end{equation}
\end{prop}
\begin{proof}
Note that $\C_u(\NS) - \C_u(S) = \fractwo \bigbig{u(\NS) - u(S)}$.
Applying twice Proposition \ref{prop:shapley_value_formula_complementary}:
\begin{equation}
    \psi_i(\C_u) = \sumSNminusi p_{S}^i \,  \bigbig{u(\NS) - u(S)} =  \psi_i(u).
\end{equation}
\end{proof}
\begin{cor}
A probabilistic value $\psi$ can be defined in terms of $\C_u$:
\begin{equation*}
        \psi_i(u) = \sumSNminusi p_{S}^i \bigbig{\C_u(\NS) - \C_u(S)}.
\end{equation*}
In particular, we have shown that each probabilistic value is associated consistent with the game map $\C$, according to definition \ref{def:associated_consistency_game_map}.
\end{cor}
\begin{proof}
Follow by $\C_u(\NS) - \C_u(S) = u(\NS) - u(S)$.
\end{proof}
\begin{prop}\label{prop:C_shapley_of_C_shapley}
\begin{equation}\label{eq:C_shapley_of_C_shapley}
    \forall S \subseteq N, \forall u \in \gameN, \quad \C_{\C_u}(S) = \, \C_u(S).
\end{equation}
\end{prop}
\begin{proof}
\begin{align*}
    \C_{\C_u}(S) &= \fractwo \, \bigbig{\C_u(S) - \C_{u}(N \setminus S) + \C_u(S)} = \C_u(S)
\end{align*}
\end{proof}

\begin{prop}\label{prop:image_game_c_u_superadditive}
$\C_u$ is superadditive if and only if, for each partition $\{R, S, T\}$ of the set of players $N$,
\begin{equation} \label{eq:condition_image_game_superadditive}
    u(R \cup S \cup T) - u(R \cup S) - u(R \cup T) - u(S \cup T) + u(R) + u(S) + u(T) \leq 0.
\end{equation}
\end{prop}
\begin{proof}
$\C_u$ is a superadditive game if and only if for all disjoint sets $S,T \subseteq N$, 
$$\C_u(S \cup T) \geq \C_u(S) + \C_u(T).$$
If we call $R = N \setminus (S \cup T)$, then $\{R, S, T\}$ forms a partition of $N$. Equation \ref{eq:condition_image_game_superadditive} follows by expanding the inequality and substituting $R$.
\end{proof}
\begin{cor}
Given a superadditive and constant sum game $u$, then $\C_u$ is superadditive.
\end{cor}
\begin{proof}
Consider a partition $\{R, S, T\}$ of $N$. By $CS_S$, it is possible to write:
$u(R \cup S \cup T) = u(R) + u(S \cup T)$.
Then, the inequality \eqref{eq:condition_image_game_superadditive} becomes:
\begin{align*}
 u(R \cup T) - u(R) - u(T) + u(R \cup S) - u(R) - u(S) \geq 0.
\end{align*}
The claim follows applying superadditivity of $u$ to the pairs $R, T$ and $R, S$. \end{proof}

\begin{cor}
If $\C_u$ is superadditive, then $\C_{\C_u}$ is superadditive.
\end{cor}
\begin{proof}
    Follows from Proposition \ref{prop:C_shapley_of_C_shapley}.
\end{proof}
\begin{prop}\label{prop:C_shapley_cohesive_constant_sum}
If $u$ is a cohesive game then 
\begin{equation}\label{eq:cohesive_C_shapley}
    \C_u(S) \geq \, u(S) \quad \forall S \subseteq N.
\end{equation}
or, equivalently, $u^*(S) \geq u$ for each coalition $S$.
The above inequalities are equalities if and only if $u$ is also constant sum.
\end{prop}
\begin{proof}
Given a cohesive game $u$ and considering the partition $\{N\setminus S, S\}$ of $N$, $u(N) \geq u(N \setminus S) + u(S)$.
Hence,
\begin{align*}
        \C_u(S) \geq \fractwo \bigbig{(u(N \setminus S) + u(S)) - u(N \setminus S) + u(S)} = u(S).
\end{align*}
Assuming that the above is an equality for each coalition, then, for each $S \subseteq N$, $u(N) - u(N \setminus S) + u(S) = 2 \, u(S)$, so $u$ is constant sum.
Vice versa, if $u$ is constant sum, then the inequality of \eqref{eq:cohesive_C_shapley} becomes an equality.
\end{proof}
In terms of probabilistic generalized values, the above proposition shows that $\C$ satisfies group rationality \cite{flores2019evaluating}.

Let us consider a game $u \in \mathcal{G}^{\{1,2\}}$. 
Then, the Shapley value is:
\begin{align*}
    \phi_1(u) &= \fractwo \bigbig{u(\{1, 2\}) - u(2) + u(1)} \\
    \phi_2(u) &= \fractwo \bigbig{u(\{1, 2\}) - u(1) + u(2)}.
\end{align*}
Note that $\phi_i(u) = \C_u(\{i\})$ for players $i = 1, 2$.
This observation extends from two-player games to quotient games by the following.
\begin{prop}\label{prop:C_shapley_and_quotient_game_2_players}
Given a game $u$, for each $S 
\subseteq N$, consider $\mathfrak{P} = \{S, \NS\}$ partition of $N$, and the quotient game $u^{\mathfrak{P}}$ Assuming that player $1$ corresponds to $S$ and player $2$ to $\NS$, then:
\begin{align}
    \C_u(S) = \phi_1(u^{\mathfrak{P}});  \qquad \C_u(\NS) = \phi_2(u^{\mathfrak{P}}).
\end{align}
\end{prop}
\begin{proof}
\begin{align*}
    \phi_1(u^{\mathfrak{P}}) &= \fractwo \bigbig{u(N) - u(N \setminus S) + u(S)} = \C_u(S). \nonumber
\end{align*}
The same computation holds for $\phi_2(u^{\mathfrak{P}})$.
\end{proof}
We suggest three meaningful interpretation of the above proposition. 
First of all, from Proposition \ref{prop:C_shapley_and_quotient_game_2_players}, the axiom CS for $\C_u$-Shapley explains as the efficiency axiom for player of the Shapley value translated to the quotient game $u^{\mathfrak{P}}$:
\begin{align*}
\C_S(u) + \C_{\NS}(u) =  \bigbig{\phi_1(u^{\mathfrak{P}}) + \phi_2(u^{\mathfrak{P}})} = u^{\mathfrak{P}}(N) = \C_N(u).
\end{align*}
Second, recall the interpretation of eq. \eqref{eq:hamiache_associated_game} provided in \cite{hamiache2001associated}: it describes the corresponding associated game $\eta^t_S$ viewing $S$ as the center of a star-like graph, and each player in $\NS$ as an isolated element. 
Instead, in the associated game $\C_u$, $S$ looks at $\NS$ as an integral entity with whom sharing the worth of the grand coalition.
Third, in terms of generalized values, note that the Shapley group value \cite{flores2019evaluating} views  $S$ as a singleton in the quotient game of $n-s+1$ players, composed by $[S] \cup \{i\in \NS\}$.
Instead, the $\C$ generalized value considers a quotient game of only 2 players, $S$ and $\NS$.

For the following, it is interesting to define a new game map that results in a scaling of the $\C$ by the factor $\fractwoNminusone$.

\begin{defin}\label{defin:Cprime}
For each game $u$ and coalition $S$, define
\begin{equation}
    \Cprime_u(S) = \fractwoNminusone \C_u(S) = \fractwoN \bigbig{u(N) - u(N \setminus S) + u(S)}.
\end{equation}
\end{defin}
By linearity, all the properties discussed for $\C$  of the propositions \ref{prop:phi_i_of_chi_shapley}, \ref{prop:C_shapley_of_C_shapley}, \ref{prop:C_shapley_cohesive_constant_sum}, and \ref{prop:C_shapley_and_quotient_game_2_players} holds also for $\Cprime$, with the corresponding equations scaled by a factor of $\fractwoNminusone$.
Clearly, the game map $\Cprime$ satisfies all the axioms listed in Definition \ref{defin:axioms_game_map}, except AvEFF.
Instead, $\Cprime$ satisfies the following efficiency axiom:
\begin{defin}\label{defin:axiom_aveff_S}
$\gamma \in \GammaN$ satisfies Efficiency (EFF) if 
\begin{equation}
    \forall u \in \gameN, \quad \sumSN \gamma_u(S) = u(N)
\end{equation} 
\end{defin}

\section{Hodge characterization}
This section presents the general mathematical framework of the combinatorial Hodge decomposition from graph theory, used in a recent characterization of the Shapley value \cite{STERN2019186} and further extended in \cite{lim2021hodge}.
Then we describe the game map $\Cprime$ as the solution of the Poisson equation on the Hasse diagram describing the game.
\subsection{Interlude: Hodge Decomposition of a graph}\label{subsec:hodge_theory_general}
\begin{notaz}
Let $V$ be a set of vertices and $E\subseteq V\times V$ be a set of edges connecting a pair of nodes in $V$.
\end{notaz}
\begin{defin}\label{defin:oriented_graph}
$G = (V, E)$ is called an oriented graph if $(a,b)\in E$ implies $(b,a)\notin E$.
\end{defin}

\begin{defin}\label{defin:elledueV_elledueE}
Denote by $\elledueV$ the space of functions $u \colon V \to \R$, %
equipped with the inner product
\begin{equation}
    \innerprod{u}{v}_{V} := \sum_{a\in V}u(a)v(a).
\end{equation}
Analogously, denote by $\elledueE$ the space of functions $f \colon E \to \R$,
equipped with the inner product
\begin{equation}
    \innerprod{f}{g}_{E} := \sum_{(a,b) \in E} f(a,b)g(a,b).
\end{equation}

\end{defin}

\begin{defin}[$\ddiff$ and $\ddstar$]
Let the graph differential be  the linear mapping $\ddiff_G \colon \elledueV \to \elledueE$ defined by
\begin{equation}\label{eq:operator_d}
    \ddiff_G \, u(a,b) := u(b) - u(a).
\end{equation}

Its adjoint $\ddstar_G$ is the unique linear map $\ddstar_G \colon \mathit{l}^2(E)\to\elledueV$ such that 
\begin{equation*}
  \innerprod{u}{\ddstar_G \, f}_{V} = \innerprod{\ddiff_G \, u}{f}_{\mathit{E}}
\end{equation*}
\end{defin}
\begin{notaz}
Whenever the graph $G$ is clear from the context, the notation omits the subscript $G$ from the differential or its adjoint, that is, $\ddiff:= \ddiff_G$ or $\ddstar:= \ddstar_G$.
\end{notaz}

\begin{defin}\label{defin:graph_laplacian}
Given a graph $G$, the graph Laplacian $L_G$ is the linear mapping $\elledueV\to\elledueV$ defined as  $L_G = \ddstar_G \, \ddiff_G$.
If clear from the context, the subscript $G$ is dropped for the graph Laplacian and we write $L$ for $L_G$.
\end{defin}

\begin{prop}\label{defin:combinatiorial_hodge_decomposition}
The inner-product spaces $\elledueV$ and $\mathit{l}^2(E)$ are decomposed as
\begin{equation}\label{eq:Hodge_Decomposition}
    \elledueV=\Ima \ddstar \oplus \Ker \ddiff \hspace{0.5cm} \mbox{and} \hspace{0.5cm} \mathit{l}^2(E) = \Ima \ddiff \oplus \Ker \ddstar,
    \end{equation}
where $\mathcal{\Ima}$ and $\Ker$ denote, as usual, the image and the kernel of a linear mapping, respectively. 
The decomposition obtained in \eqref{eq:Hodge_Decomposition} is referred to as the combinatorial Hodge decomposition of $\elledueV$ and $\mathit{l}^2(E)$.
\end{prop}
\begin{proof}
Although this is a well-known result, we report here a simple proof of it that enlightens some aspects of this decomposition which will be of use afterward.

Given any pair of inner product linear spaces $V$ and $W$ and a linear mapping $f:V\to W$ it holds that 
\[ f^*(w)=0_V \iff 0=\innerprod{v}{f^*(v)}_{V}=\innerprod{f(v)}{w}_{W},\, \forall v\in V \] 
so that $\Ker f^*=\{w\in W \,: \innerprod{f(v)}{w}_{W}=0,\, \forall v\in V\}$, that is, $\Ker f^*=(\Ima f)^{\perp},$ the orthogonal complement of $\Ima f$ in $W$. 

\textit{Mutatis mutandis} and using $f=(f^*)^*$ it follows that $\Ker f=(\Ima{f^*})^{\perp}$ as well. 
Therefore, 
\begin{equation}
    V=\Ker{f}\oplus \Ima{f^*}=\Ker{f}\oplus(\Ker{f})^{\perp}
\end{equation}
and 
\begin{equation}
     W=\Ima f\oplus \Ker{f^*}=\Ima f\oplus(\Ima f)^{\perp}
\end{equation}
and the results follows by specializing to $\ddiff:\elledueV\to \mathit{l}^2(E)$
\end{proof}
\subsection{Poisson Equation}\label{subsec:poiss_equat}
\begin{thm} \label{theo:poisson}
Let $G=(V, E)$ be a finite simple directed graph and let $H=(V,F)$ be a sub-graph of $G$ with $F\subset E$. Then, for all $v\in\elledueV$, the equation
\begin{equation}\label{pussy}
    L_G\,x=L_H\,v
\end{equation}
has a solution $x=v_H\in \elledueV.$
\end{thm}

\begin{proof}
To prove the statement we need to introduce the linear mapping $\iota_F:\mathit{l}^2(F)\to \mathit{l}^2(E)$ that is induced by the inclusion $F \subset E$, namely 
\begin{equation}
\iota_F(\psi)(a,b):=\begin{cases}
 \psi(a,b) & \text{if } (a,b)\in F\\
 0 & \text{otherwise}
\end{cases}       
\end{equation}
for all $\psi\in \mathit{l}^2(F)$ and $(a,b)\in E.$
As it is easy to check, the following holds $\iota_F^* \iota_F = id_{\mathit{l}^2(F)},$ where $\iota_F^*$ is the adjoint of $\iota_F$ with respect to $\innerprod{\,}{\,}_{E}$ and $\innerprod{\,}{\,}_{F}.$

Now observe that $\dds{H}=\iota_F^*\dds{G}$, so that $L_Hv=\dds{H}^*\dds{H}v=\ddstar_G\iota_F\dds{H}v$. Because $\iota_F\dds{H}v\in\elledueE$, there are $v_H \in \elledueV$ and $r_H(v)\in \Ker\ddstar_G$ such that $\iota_F\dds{H}v=\ddiff_G v_H+r_H(v)$. 
We can conclude that:
\begin{equation}
L_H \, v = \dds{H}^*\dds{H} \, v= \ddstar_G \iota_F \dds{H} \, v = \ddstar_G (\ddiff_G \, v_H + r_H(v) ) = \ddstar_G\, \ddiff_G\,v_H = L_G v_H
\end{equation}
\end{proof}
The following is a well-known fact we report here for the readers' comfort.
\begin{prop}
A graph is connected if and only if the kernel of its Laplacian is one-dimensional.
\end{prop}
\begin{proof}
 First, observe that $L=\ddstar\ddiff$ implies that $\Ker L = \Ker\ddiff$, and $\Ker \ddiff = \{u \in \elledueV \, \mid \, u(x) = u(y), \, \forall (x,y) \in E\}.$ 
 Let $C_1,\dots, C_{\beta}$ be the connected components of $G$ and let $\mathbb{1}_i$ be the characteristic function of $C_i$ for $i=1,\dots, \beta$, then $(\mathbb{1}_1,\dots, \mathbb{1}_{\beta})$ is a basis of $\Ker d$.
It follows that $G$ is connected if and only if $\Ker \ddiff =\{\alpha\mathbb{1}\,:\, \alpha\in\mathbb{R}\}$, where $\mathbb{1}(x)=1$ for all $x\in V$.
\end{proof}
From now on, unless otherwise stated, we assume the graph $G$ to be connected. 

\begin{prop}\label{prop:uniquesol}
For all $\alpha\in V$ denote by $V_{\alpha}$ the subspace $V_{\alpha}=\{u\in\elledueV\,:\, u(\alpha)=0\}$ of $\elledueV.$ Then, for all $u\in\elledueV$, and for all $H=(V,F)$ with $F\subset E$, the Poisson Equation (\ref{pussy}) has a unique solution $u_H^{\alpha}\in V_{\alpha}$.
\end{prop}
\begin{proof}
Let $u_H\,\in\,\elledueV$ be such that $L_H\,u=L_G\,u_H$, then $u_H-u_H(x)\mathbb{1}$ is another solution of (\ref{pussy}) that belongs to $V_{\alpha}$.
Assume then that $u_H^{\alpha}, v_H^{\alpha} \in V_{\alpha}$ are two solutions of (\ref{pussy}), then $u_H^{\alpha} - v_H^{\alpha}$ is constant on $V$ and belongs to $V_{\alpha}$, therefore it is null and $u_H^{\alpha} = v_H^{\alpha}$.
\end{proof}

We now focus on a family of graphs related to TU games, Shapley value, and $\C$.
We first need a definition.
\begin{defin}\label{defin: subsetgraph} Given the usual set of players $N$, of cardinality $n\in \mathbb{N}$, we will denote with $V$ a set of vertices in bijection with $\powerset{N}.$ We set
\begin{enumerate}
\item $G :=(V,E),$  $E = \{(A,B)\in V\times V \,:\, A\subset B\};$
\item $G^k :=(V,E^k),$  $E^k = \{(A,B)\in E \, : \, |B\setminus A|\leq k\},$ for all $1\leq k \leq n;$
\item $G_S := (V, E_S),$  $E_S = \{(A,B)\in E \, : \, B\setminus A = S \},$ for all $S\in\powerset{N}.$   
\end{enumerate}
Accordingly we will denote by $\ddiff,$ $\ddiff^k,$ $\ddiff_S,$ $L,$ $L^k$ and $L_S$ the corresponding differentials and Laplacians.
\end{defin}

Then we can formulate the following statement: a direct corollary of Th.\ref{theo:poisson} and Prop.\ref{prop:uniquesol}.

\begin{cor}\label{cor:poissonk}
Let $S\in\powerset{N}$, $u\in\gameN$ and $k$ an integer $|S|\leq k \leq n$. Then, the equation 
\begin{equation}\label{eq:pussyk}
L^kx=L_Su     
\end{equation}
admits a unique solution $x=u_S^k\in\gameN$.
\end{cor}
\begin{proof}
It follows from Th.\ref{theo:poisson} and Prop.\ref{prop:uniquesol} by setting $G=G^k$, $H=G_S$ and $V_{\alpha}=V_{\emptyset}=\gameN$.
   \end{proof}

\begin{defin}\label{def:sstarzero_sstarone}
For all permutations of the $N$ players $\sigma \in \mathcal{S}_N$, define $\sigma^* \colon \elledueV \to \elledueV$ by $(\sigma^* u)(S) := u \bigbig{\sigma (S)}$, for all $u\in \elledueV$ and $S\subset N$. 
\end{defin}
\begin{rmk}
For all $\sigma \in \mcal{S}_N$ and $A\subseteq B\subseteq N$ it holds $\sigma (A\setminus B)=\sigma(A)\setminus \sigma(B)$.
\end{rmk}

\bigskip

\begin{thm}\label{thm:decomposition_S_differential_game}
    Let $S\in\powerset{N}$, $|S|\leq k\leq n$, $G$, $G^k$ and $G_S$ be as in \ref{defin: subsetgraph}. Let $u_S^k$ be the unique game in $\gameN$ that is a solution of the equation (\ref{eq:pussyk}) for $u\in\gameN$ as in Corollary\ref{cor:poissonk}.
    
    Then, the games $u_S^k$ satisfy the following:
    \begin{itemize}
        \item[(a)] $\sum_{S:|S|\leq k} \, u_S^k = u$;
        \item[(b)] if $u(S \cup T) = u(T)$ for all $T \subseteq N \setminus S$, then $u_S^k=0$;
        \item[(c)] if $\sigma \in \mcal{S}_N$, then $(\sigma^*u)_S^k=\sigma^*\,(u_{\sigma(S)}^k)$. 
        \item[(d)] For any two games $u,v$ and $\alpha,
        \beta\in\mathbb{R}$, then $(\alpha\, u+\beta\,v)_S^k=\alpha \,u_S^k+\beta \, v_S^k$
    \end{itemize}
\end{thm}
\begin{proof} 

(a) Let $S\in\powerset{N}$, then $\sum_{S\in\powerset{N}}\iota_S\iota_S^*=id_{\mathit{l}^2(E)}$ as it is easy to check. Therefore,
$\sum_{S\in\powerset{N}}L_S=\sum_{S\in\powerset{N}}d^*\iota_S\iota_S^*d=d(\sum_{S\in\powerset{N}}\iota_S\iota_S^*)d^*=dd^*=L$. Henceforth 
$L\sum_{S\in\powerset{N}}u_S=\sum_{S\in\powerset{N}}Lu_S=\sum_{S\in\powerset{N}}L_Su=Lu$ and the thesis follows because $\ker(L)\cap \gameN=\{\mathbf{0}\}$.

(b) It is enough to observe that
\begin{equation}\label{eqn: lsut}
 L_Su(T)=
\begin{cases}
    u(T)-u(T\cup S) & \text{if $T\cap S=\emptyset$} \\
    u(T)-u(T\setminus S) & \text{if $T\cap S=S$} \\
    0 & \text{otherwise}
\end{cases}
\end{equation}
so that, if $S$ is null then $L_Su=0$ and, henceforth, $Lu_S=0$. Then $u_S=0$ since $\ker L\cap \gameN=\{0\}.$

(c) The first observation is that $L\sigma^*u=\sigma^*Lu$. Indeed, for all $A\in\powerset{N}$ and $u\in\GammaN$, 
$L\sigma^*u(A)=\deg(A)\sigma^*u(A)-\sum_{B\sim A}\sigma^*u(B)=\sigma^*Lu(A)$ where $\deg(\sigma(A))=\deg(A)$ because $|A|=|\sigma(A)|$.
The second observation is that $L_S\sigma^*u=\sigma^*L_{\sigma(S)}u$ as it follows directly from the definition of $\sigma^*$ and the equation (\ref{eqn: lsut}).
Then, $L(\sigma^*u)_S=L_S\sigma^*u=\sigma^*L_{\sigma(S)}u=\sigma^*Lu_{\sigma(S)}=L\sigma^*(u_{\sigma(S)})$ and the result follows by the injectivity of the Laplacian $L$ on $\gameN$. 
(d) Trivial.
\end{proof}

There is a deep connection between the Hodge-Shapley associated game and the solution of the Poisson equation which is enlightened in the theorem below.
\begin{thm}\label{thm:ucprime}
Let $S\in\powerset{N}$, $G$, and $G_S$ be as in \ref{defin: subsetgraph}. Let $u_S$ be the unique game in $\gameN$ that is a solution of the Poisson equation (\ref{pussy}) for $u\in\gameN$ and $H=G_S$. Then $u_S(N)=\Cprime_u(S).$
\end{thm}

\begin{proof}
First recall that, for a generic graph $G=(V,E)$, $v\in V$, and $u \in \elledueV$ it holds that $L\,u(v) = \deg_G(v)\,u(v) - \sum_{w\sim v}u(w)$, where $w\sim v$ if and only if $(v,w)\in E$ or $(w,v)\in E$.
Then, recall that $u_S\in\gameN$ so that $u_S(\emptyset)=0$. 
It follows that
\begin{equation}
\begin{split}
Lu_S(\emptyset) & = (2^n-1) \, u_S(\emptyset)-\sum_{A\neq \emptyset}u_S(A)\\ 
& =2^nu_S(\emptyset)-\sum_{A\in V}u_S(A)\\
& =-\sum_{A\in V}u_S(A) 
\end{split}
\end{equation}
and 
\begin{equation}
    L_S\,u(\emptyset) = u(\emptyset)-u(S) = -u(S).
\end{equation}
Therefore, 
\begin{equation}\label{eq: ussum}
u(S)=\sum_{A\in V}u_S(A).    
\end{equation}

On the other hand
\begin{equation}\label{eq: lusn}
\begin{split}
L\,u_S(N) & = (2^n-1)u_S(N)-\sum_{A\neq N}u_S(A)\\
& = 2^nu_S(N)-\sum_{A\in V}u_S(N)
\end{split}
\end{equation}
and 
\begin{equation}\label{eq: lsun}
    L_S\,u(N) = u(N) - u(\NS)
\end{equation}

Consider now that $L\,u_S(N)=L_S\,u(N),$ so that, by using Eq.\eqref{eq: ussum}, \eqref{eq: lusn} and \eqref{eq: lsun} we obtain that 
\begin{equation}
    u_S(N)=\fractwoN(u(N) - u(\NS)+u(S))
\end{equation}
and the thesis follows.
\end{proof}

At the opposite extreme of the scale, when $k=1$, the Poisson equation's solution provides the Shapley value as first proven (although stated in a slightly different manner) in \cite{STERN2019186}.
\begin{thm}
    Let $1\leq i \leq n$ and, given $u\in\gameN$, let $u^1_i$ be the unique game that is the solution of the Poisson Equation $L^1u^1_i=L_iu$. Then $u^1_i(N)$ is the $i-$th Shapley value of $u$, that is $u^1_i(N)=\phi_i(u).$
\end{thm}

\subsection{Examples}
\begin{ex}
Consider the classic glove game \cite{STERN2019186}, described as $u \in \gameN$, with $n=3$. 
Then we can compare $u$ and $\C_u$ (dropping set parenthesis, e.g. $ijk = \{i,j,k\})$:
\begin{align*}
    u(S) = \begin{cases}
    1 \text{ if } S = 12, 13, 123 \\
    0 \text{ otherwise}.
    \end{cases} 
    \qquad
    \C_u(S) = \begin{cases}
    1 \text{ if } S = 12, 13, 123 \\
    \frac{1}{2} \text{ if } S = 1, 23 \\
    0 \text{ otherwise}.
    \end{cases}
\end{align*}
We can see that the only difference resides in the payoff of coalitions $1$ and $23$.
\end{ex}
\begin{ex} 
Bankruptcy problems \cite{2010_gonzalez_introductory} hold back to the Talmud: a man dies leaving an estate of $E=200$, and three creditors ask for compensation of $c_1 = 100$, $c_2 = 200$, and $c_3 = 300$, respectively. 
A formulation in terms of a TU-game with $n=3$ players is due to \cite{oneill1982problem}.
In this game, $u(S) = \max \bigbig{0, E - \sum_{i \in \NS} c_i}$. 
In particular, the computation of $\C_u$ yields 
\begin{align*}
    u(S) = \begin{cases}
        200 \text{ if } S=123 \\
        100 \text{ if } S=23 \\
        0 \text{ otherwise}.
    \end{cases} \qquad \C_u(S) = \begin{cases}
        200 \text{ if } S=123 \\
        150 \text{ if } S=23 \\
        50 \text{ if } S = 1 \\
        0 \text{ if } S = \emptyset \\
        100 \text{ otherwise}.
    \end{cases}
\end{align*}

Note that $u$ is a superadditive game while $\C_u$ is not (just compare $\C_u$ on singletons $1,2,3$ with $\C_u(N)$).
In addition, each pair of complementary coalitions is constant sum, and, in particular non-trivial pairs, except $(1, \{2,3\})$ shares half of the grand coalition payoff.
$\C_{1}$ is lower than the others because of the role of $\{2,3\}$ in the game $u$.   
\end{ex}
\begin{ex}
Airport problems deal with the redistribution of cost among movements of aircraft on a runaway \cite{2010_gonzalez_introductory}. 
Aircraft of different kinds need different lengths of runaway.
Let us assume that there are $4$ movements represented as players.
The first movement costs $12$, the second and third movements cost $28$ each, and the fourth movement costs $30$. 
The characteristic function $u \in \gameN$, $n=4$, is described as follows:
\begin{equation*}
    u(S) = \begin{cases}
        -12 \text{ if } S = 1 \\
        -28 \text{ if } S \cap 2,3 \neq \emptyset \text{ and } 4\notin S\\
        -30 \text{ otherwise.}
    \end{cases}
\end{equation*}
By computing $\C_u$ we obtain:
\begin{equation*}
    \C_u(S) = \begin{cases}
        -6 \text{ if } S = 1 \\
        -24 \text{ if } S = 234\\
        -14 \text{ if } S \cap 23 \neq \emptyset \text{ and } 4\notin S \\
        -16 \text{ otherwise.}\\
    \end{cases}
\end{equation*}
Note that both $u$ and $\C_u$ are superadditive. 
\end{ex}

\section{Conclusions}
In this work, we proposed a novel view on the study of solution concepts of TU-games. 
In particular, we suggested to unifying generalized values and associated games with the definition of game maps.
We defined a novel game map $\C$ and its corresponding associated game $\C_u$, and we provided an axiomatic characterization inspired by generalized values.

The game map $\C$ has significant properties, that we can summarize as follows.
First, all probabilistic values are associated consistent with $\C$.
Second, it can be interpreted as a sharing rule to allocate the grand coalition payoff between each pair of complementary coalitions in the original game.
In particular, the sharing follows a fairness principle, as expressed by the bilaterality, constant sum, and null coalition axioms.
Therefore, the associated game $\C_u$ can be deemed as a fair version of the original game.
Third, $\C$ follows the equal division of surplus principle for the allocation of the grand coalition payoff between complementary coalitions.
In this sense, it appears a strong link between $\C$ and the Shapley value in the quotient game of two players.
Fourth, as a probabilistic generalized value, it computes the influence or power of complementary coalitions, that is group rational for cohesive games.

Furthermore, we showed that by multiplying $\C$ with a scaling factor, we obtain a game map $\Cprime$ satisfying the efficiency axiom for coalitions and retaining the properties of $\C$.
Finally, we introduced the transitive closure of the Hasse diagram graph representing the inclusion relation between coalitions in a game.
Surprisingly, the solution of the Poisson equation derived from the graph uniquely characterizes the game map $\Cprime$.

\section*{Acknowledgement} 
This study was carried out within the FAIR - Future Artificial Intelligence Research and received funding from the European Union Next-GenerationEU (PIANO NAZIONALE DI RIPRESA E RESILIENZA (PNRR) – MISSIONE 4 COMPONENTE 2, INVESTIMENTO 1.3 – D.D. 1555 11/10/2022, PE00000013). This manuscript reflects only the authors’ views and opinions, neither the European Union nor the European Commission can be considered responsible for them.
A.M. deeply thanks Mauro Leoncini for his kind support.

\printbibliography

\end{document}